\documentclass[11pt,dvips]{article}
\usepackage{eurosym}
\usepackage{graphicx}
\usepackage{amsmath}
\usepackage{amssymb}
\usepackage{latexsym}
\usepackage{color}
\usepackage{epstopdf}

\input{tcilatex}
\begin{document}

\begin{center}
\vspace{0.7cm}

{\Large \textbf{Gaussian quantum steering in a nondegenerate three-level
laser}}

\vspace{0.7cm}

\textbf{B. Boukhris}$^{a,b}${\footnote{%
email: \textsf{b.boukhris@uiz.ac.ma}}}, \textbf{A. Tirbiyine}$^{a}${%
\footnote{%
email: \textsf{a.tirbiyine@uiz.ac.ma}}}, and \textbf{J. El Qars}$^{a}${%
\footnote{%
email: \textsf{j.elqars@uiz.ac.ma}}}

\vspace{0.5cm}

$^{a}$\textit{LMS3E, Faculty of Applied Sciences, Ibn Zohr University,
Agadir, Morocco}

$^{b}$\textit{LASIME, National School of Applied Sciences, Ibn Zohr
University, Agadir, Morocco}

\vspace{0.9cm}\textbf{Abstract}
\end{center}

Steering is a type of nonseparable quantum correlation, where its inherent
asymmetric feature makes it distinct from Bell-nonlocality and entanglement.
In this paper, we investigate quantum steering in a two-mode Gaussian state $%
\hat{\varrho}_{c_{1}c_{2}}$ coupled to a two-mode vacuum reservoir. The mode
$c_{1}$($c_{2}$) is emitted during the first(second) transition of a
nondegenerate three-level cascade laser. By means of the master equation of
the state $\hat{\varrho}_{c_{1}c_{2}}$, we derive analytical expression of
the steady-state covariance matrix of the modes $c_{1}$ and $c_{2}$. Using
realistic experimental parameters, we show that the state $\hat{\varrho}%
_{c_{1}c_{2}}$ can exhibit asymmetric steering. Furthermore, by an
appropriate choice of the physical parameters of the state $\hat{\varrho}%
_{c_{1}c_{2}}$, we show that one-way steering can be achieved. Essentially,
we demonstrate that one-way steering can, in general, occur only from $%
c_{1}\rightarrow c_{2}$. Besides, we perform a comparative study between the
steering of the two laser modes and their Gaussian R\'{e}nyi-2 entanglement.
As results, we found that the entanglement and steering behave similarly in
the same circumstances, i.e., both of them decay under dissipation effect,
moreover, they can be well enhanced by inducing more and more quantum coherence in the state $\hat{\varrho}_{c_{1}c_{2}}$%
. In particular, we found that the steering remains always less than the
Gaussian R\'{e}nyi-2 entanglement.

\section{Introduction}

The aspect of quantum steering was first introduced by E. Schr\"{o}dinger
\cite{EPR1,EPR2} in order to capture the essence of the
Einstein-Podolsky-Rosen paradox \cite{EPR}. According to E. Schr\"{o}dinger,
steering is a quantum phenomenon that allows one observer (say Alice) to
remotely affect the state of another observer (say Bob) via local operations
\cite{wiseman}.

In the hierarchy of nonseparable quantum correlations \cite{wiseman},
steering stands between entanglement \cite{ENT} and Bell nonlocality \cite%
{Bell}. From an operational point of view, in a bipartite quantum state $%
\hat{\varrho}_{XY}$, violation of the Bell inequality implies that the state
is steerable in both directions $X\rightleftarrows Y$, while, demonstrating
steering at least in one direction ($X\rightarrow Y$) or ($Y\rightarrow X$)
makes sure that the state is entangled.

In quantum information precessing \cite{Nielson}, quantum steering
corresponds to the task of verifiable entanglement distribution by an
untrusted observer \cite{wiseman}. In other words, if two observers Alice
and Bob share a bipartite state which is steerable (saying) from Alice$%
\rightarrow $Bob, therefore, Alice can convince Bob that their shared
bipartite state is entangled by implementing local measurements and
classical communications \cite{wiseman}. Unlike entanglement, quantum
steering is an asymmetric form of quantum correlations, i.e., a bipartite
state $\hat{\varrho}_{XY}$ may be steerable solely in one direction, which
is referred to as one-way quantum steering \cite{kogias}.

To probe quantum steering in bipartite quantum states, various criteria have
been proposed. We cite for instance, the Reid criterion \cite{Reid}, the
Cavalcanti criterion \cite{Cavalcanti}, the Costa criterion \cite{Costa},
and the Wolmann criterion \cite{Walborn}. In particular, within the Gaussian
framework \cite{GS}, Kogias and the co-authors \cite{kogias} have proposed a
computable measure to quantify the amount by which a bipartite state is
steerable in a given direction under Gaussian measurements. Also, they
showed that in an arbitrary two-mode Gaussian state, the proposed measure
cannot exceed entanglement when the later is defined via the Gaussian R\'{e}%
nyi-2 entropy \cite{R1,R2}.

Nowadays, it is believed that the key ingredient of asymmetric quantum
information tasks---where some of the parties are untrusted---is quantum
steering \cite{Uola}. For example, quantum steering has been recognized as
the essential resource for one-sided device-independent (1SDI) quantum key
distribution \cite{1SDI}, quantum secret sharing \cite{QSC}, quantum
teleportation \cite{QTP}, one-way quantum computing \cite{OWQC}, and
sub-channel discrimination \cite{SCD}.

Gaussian quantum steering has been studied in various systems \cite%
{O1,O2,O3,M1,M2,M3,GCS,CST,Olsen,Ming,cel1,cel2}. To the best of our
knowledge, no previous work considering a nondegenerate three-level laser
coupled to two-mode vacuum reservoir has analyzed asymmetric Gaussian
steering.

Motivated by the considerable attention that has recently been paid to
quantum steering as the key ingredient in the implementation of asymmetric
quantum information tasks, we investigate, under realistic experimental
conditions, Gaussian quantum steering of two optical modes (labelled as $%
c_{1}$ and $c_{2}$) generated by a nondegenerate three-level cascade laser.
For this, we use the measure proposed by Kogias \textit{et al.} \cite{kogias}
to quantify the amount by which the two-mode Gaussian state $\hat{\varrho}%
_{c_{1}c_{2}}$ is steerable in a given direction. Furthermore, we compare,
under the same circumstances, the steering of the two considered modes with
their corresponding entanglement quantified by means of the Gaussian R\'{e}%
nyi-2 entanglement. We emphasize that our interest in Gaussian states is
motivated on the one hand by the fact that such states admit a simple
mathematical description, and on the other hand by the fact that they can be
reliably generated and manipulated in a variety of experimental platforms
\cite{GS}. Importantly, Gaussian measurements can be effectively implemented
by means of homodyne detections \cite{Laurat}.

Extensive researches have been carried out on the quantum analysis of the
light generated by a nondegenerate three-level laser \cite{S and Z}. These
researches reveal that the two laser modes, emitted during a cascade
transition, exhibit strong non-classical correlations between them \cite{Han}%
, which is exploited to study, e.g., entanglement, quantum bistability \cite%
{sete1}, squeezing \cite{sete2} and the statistical properties of the light
\cite{Fesseha}.

A nondegenerate three-level cascade laser can be defined as a quantum system
constituted by a set of nondegenerate three-level atoms, which are initially
prepared in a quantum coherent superposition of the top and bottom levels
inside a cavity \cite{Tesfa}.

In such system, the fundamental role is played by the quantum coherence, that can be achieved either by an
initial preparation of the atoms in a coherent quantum superposition of the
top and bottom levels \cite{Bergout}, or by coupling these two levels by
means of strong laser field \cite{Ansari2}. When a single atom passes from
the top to bottom level through the intermediate level, two photons are
created. If the two emitted photons are identical (i.e., have the same
frequency), the laser is referred to as a degenerate three-level laser, and
a nondegenerate three-level laser otherwise.

The remainder of this paper is organized as follows. In Sec. 2, we introduce
the studied system, and by applying the master equation of the two-mode
Gaussian state $\hat{\varrho}_{c_{1}c_{2}}$, we calculate the dynamics of
the first and second moments of the two laser modes variables. Also, we
analytically evaluate the steady-state covariance matrix fully describing
the two-mode Gaussian state $\hat{\varrho}_{c_{1}c_{2}}$. In Sec. 3, using
realistic experimental parameters, we study Gaussian quantum steering of the
two laser modes $c_{1}$ and $c_{2}.$ Furthermore, we compare the steering of
the modes $c_{1}$ and $c_{2}$ with their corresponding entanglement
quantified by means of the Gaussian R\'{e}nyi-2 entanglement. Finally, in
Sec. 4, we draw our conclusions.

\section{Model and master equation}

\begin{figure}[tbh]
\centerline{\includegraphics[width=10.5cm]{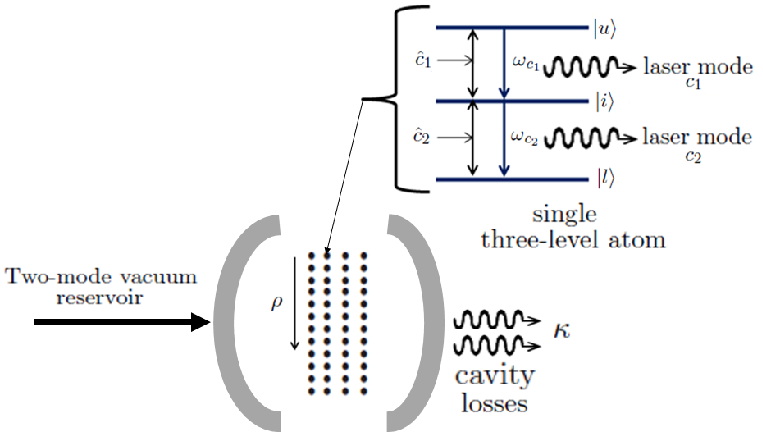}}
\caption{Schematic representation of a nondegenerate three-level laser
coupled to a two-mode vacuum reservoir \protect\cite{Blockly}. $|u\rangle$, $%
|i\rangle$, and $|l\rangle$ denote, respectively, the upper, intermediate,
and lower energy levels for a single three-level atom. The transition from $%
|u\rangle \rightarrow |i\rangle $($|i\rangle \rightarrow |l\rangle $), at
frequency $\protect\omega_{c_{1}}$($\protect\omega_{c_{2}}$) and spontaneous
emission decay rate $\protect\gamma _{ul}$($\protect\gamma _{li}$), is
considered to be in resonance with the quantized cavity modes $c_{1}$($c_{2}$%
). Whereas, the transition from $|u\rangle \rightarrow |i\rangle $ is dipole
forbidden \protect\cite{Han}. The gain medium of the laser is an ensemble of
three-level atoms in a cascade configuration. The atoms are assumed to be
initially prepared in a coherent superposition of the upper and lower
levels, and injected into the cavity with a constant rate $\protect\rho$.}
\label{f1}
\end{figure}

Inside a doubly resonant cavity, we consider a set of nondegenerate
three-level atoms in interaction with two optical modes of the quantized
cavity field \cite{S and Z}. The first(second) optical mode is specified by
its annihilation operator ($\hat{c}_{1}$)$\hat{c}_{2}$, frequency ($\omega
_{c_{1}}$)$\omega _{c_{2}}$ and dissipation rate ($\kappa _{1}$)$\kappa _{2}$%
. We assume that the atoms are injected into the cavity with a constant rate
$\rho$ and removed during a time $\tau $ \cite{Fesseha}. As illustrated in
Fig. \ref{f1}, the upper, intermediate and lower energy levels of a single
nondegenerate three-level atom are denoted, respectively, by $|u\rangle ,$ $%
|i\rangle $ and $|l\rangle $.
Under the rotating wave approximation, the two optical modes $c_{1}$ and $%
c_{2}$ and a single three-level atom can be described by the following
Hamiltonian \cite{S and Z}
\begin{equation}
\mathcal{\hat{H}}_{\mathrm{int}}=\mathrm{i}\hbar \left( \epsilon _{ui}\hat{c}%
_{1}|u\rangle \langle i|+\epsilon _{il}\hat{c}_{2}|i\rangle \langle
l|-\epsilon _{ui}|i\rangle \langle u|\hat{c}_{1}^{\dag }-\epsilon
_{il}|l\rangle \langle i|\hat{c}_{2}^{\dag }\right) ,  \label{e1}
\end{equation}%
where \textrm{int} stands for interaction picture. $\epsilon _{ui}$ and $%
\epsilon _{il}$ being the coupling constants corresponding to the
transitions $|u\rangle \rightarrow |i\rangle $ and $|i\rangle \rightarrow
|l\rangle $, respectively \cite{Fesseha}. The atoms are also assumed to be
initially prepared in an arbitrary quantum coherent superposition of the
upper $|u\rangle $ and lower $|l\rangle $ energy levels \cite{CS}. With this
assumption, the initial state of a single atom as well as its associated
density operator, respectively, read
\begin{eqnarray}
|\Psi _{\mathrm{sa}}(0)\rangle &=&P_{u}|u\rangle +P_{l}|l\rangle  \label{e2}
\\
\hat{\varrho}_{\mathrm{sa}}(0) &=&\varrho _{uu}^{(0)}|u\rangle \langle
u|+\varrho _{ul}^{(0)}|u\rangle \langle l|+\varrho _{lu}^{(0)}|l\rangle
\langle u|+\varrho _{ll}^{(0)}|l\rangle \langle l|,  \label{e3}
\end{eqnarray}%
where \textrm{sa} stands for single atom. In the last equation, $\varrho
_{uu}^{(0)}=|P_{u}|^{2}$ and $\varrho _{ll}^{(0)}=|P_{l}|^{2}$ represent the
probabilities for a single three-level atom to be initially in the upper and
lower levels, respectively, while $\varrho _{ul}^{(0)}=\varrho
_{lu}^{(0)\ast }=P_{u}P_{l}^{\ast }$ is the initial quantum
coherence of a single three-level atom \cite{SEtte}. From now on, without
affecting the generality of our study, we take identical spontaneous decay
rates for both transitions $|u\rangle \rightarrow |i\rangle $ and $|i\rangle
\rightarrow |l\rangle $, i.e., $\gamma _{ui}=\gamma _{il}=\gamma $,
identical coupling transitions, i.e., $\epsilon _{ui}=\epsilon
_{il}=\epsilon $, and identical cavity decay rates $\kappa _{1}=\kappa
_{2}=\kappa .$

The master equation for the reduced density operator $\hat{\varrho}%
_{c_{1}c_{2}}\equiv \hat{\varrho}$ of the two laser modes $c_{1}$ and $c_{2}$%
, reads \cite{Louissel,Akramine}
\begin{equation}
\partial _{t}\hat{\varrho}=\frac{-\mathrm{i}}{\hbar }\mathrm{Tr}_{\mathrm{sa}%
}\left[ \mathcal{\hat{H}}_{\mathrm{int}},\hat{\varrho}_{\{\mathrm{sa+field}%
\}}\right] +\sum\limits_{j=1,2}\frac{\kappa _{j}}{2}\mathcal{L}\left[ \hat{c}%
_{j}\right] \hat{\varrho},  \label{e4}
\end{equation}%
where $\hat{\varrho}_{\{\mathrm{sa+field}\}}$ is the density matrix that
describes a single atom plus the two-mode cavity field, and the Lindblad
operator $\mathcal{L}\left[ \hat{c}_{j}\right] \hat{\varrho}=2\hat{c}_{j}%
\hat{\varrho}\hat{c}_{j}^{\dag }-\left[ \hat{c}_{j}^{\dag }\hat{c}_{j},\hat{%
\varrho}\right] _{+}$ is added to describe the damping of the $j$-\textrm{th
}laser mode in the vacuum reservoir.

In the good cavity limit $\gamma \gg \kappa $, and the linear-adiabatic
approximation \cite{sargent}, one can show---after some tedious but
straightforward algebra---that Eq. (\ref{e4}) would be \cite{elqars1,elqars2}
\begin{eqnarray}
\partial _{t}\hat{\varrho} &=&\frac{A\varrho _{uu}^{(0)}}{2}[2\hat{c}%
_{1}^{\dag }\hat{\varrho}\hat{c}_{1}-\hat{c}_{1}\hat{c}_{1}^{\dag }\hat{%
\varrho}-\hat{\varrho}\hat{c}_{1}\hat{c}_{1}^{\dag }]+\frac{A\varrho
_{ll}^{(0)}}{2}[2\hat{c}_{2}\hat{\varrho}\hat{c}_{2}^{\dag }-\hat{c}%
_{2}^{\dag }\hat{c}_{2}\hat{\varrho}-\hat{\varrho}\hat{c}_{2}^{\dag }\hat{c}%
_{2}]+  \notag \\
&&\frac{A\varrho _{ul}^{(0)}}{2}[\hat{\varrho}\hat{c}_{1}^{\dag }\hat{c}%
_{2}^{\dag }+\hat{c}_{1}^{\dag }\hat{c}_{2}^{\dag }\hat{\varrho}+\hat{c}_{1}%
\hat{c}_{2}\hat{\varrho}+\hat{\varrho}\hat{c}_{1}\hat{c}_{2}-2\hat{c}%
_{1}^{\dag }\hat{\varrho}\hat{c}_{2}^{\dag }-2\hat{c}_{2}\hat{\varrho}\hat{c}%
_{1}]+\sum\limits_{j=1,2}\frac{\kappa }{2}\mathcal{L}\left[ \hat{c}_{j}%
\right] \hat{\varrho},  \label{e5}
\end{eqnarray}%
where we choice $\varrho _{ul}^{(0)}$ to be real for simplicity. The linear
gain coefficient $A=2\rho \epsilon ^{2}/\gamma ^{2}$ describes the rate at
which the three-level atoms are injected into the cavity \cite{tesfa}. In
the above equation, the term proportional to $\varrho _{uu}^{(0)}$($\varrho
_{ll}^{(0)}$) represents the gain(loss) of the laser mode $c_{1}$($c_{2}$),
whereas the term proportional to $\varrho _{ul}^{(0)}$ represents the
coupling between these two laser modes due to the quantum
coherence \cite{S and Z}. Now, using Eq. (\ref{e5}) and the formula $%
\partial _{t}\langle \mathcal{\hat{O}}(t)\rangle =\mathrm{Tr}(\partial _{t}%
\hat{\varrho}\mathcal{\hat{O}}(t))$, we obtain the dynamics of the first and
second moments of the laser modes variables,
\begin{eqnarray}
\partial _{t}\langle \hat{c}_{j}(t)\rangle &=&\frac{-\Gamma _{j}}{2}\langle
\hat{c}_{j}(t)\rangle +\frac{\left( -1\right) ^{j}A\varrho _{ul}^{(0)}}{2}%
\langle \hat{c}_{3-j}^{\dag }(t)\rangle ,\text{ for }j=1,2\text{,}
\label{e6} \\
\partial _{t}\langle \hat{c}_{j}^{2}(t)\rangle &=&-\Gamma _{j}\langle \hat{c}%
_{j}^{2}(t)\rangle +\left( -1\right) ^{j}A\varrho _{ul}^{(0)}\langle \hat{c}%
_{1}(t)\hat{c}_{2}^{\dag }(t)\rangle ,  \label{e7} \\
\partial _{t}\langle \hat{c}_{j}^{\dag }(t)\hat{c}_{j}(t)\rangle &=&-\Gamma
_{j}\langle \hat{c}_{j}^{\dag }(t)\hat{c}_{j}(t)\rangle +\frac{\left(
-1\right) ^{j}A\varrho _{ul}^{(0)}}{2}\left[ \langle \hat{c}_{1}^{\dag }(t)%
\hat{c}_{2}^{\dag }(t)\rangle +\langle \hat{c}_{1}(t)\hat{c}_{2}(t)\rangle %
\right] +(2-j)A\varrho _{uu}^{(0)},  \label{e8} \\
\partial _{t}\langle \hat{c}_{1}(t)\hat{c}_{2}(t)\rangle &=&-\frac{\Gamma
_{1}+\Gamma _{2}}{2}\langle \hat{c}_{1}(t)\hat{c}_{2}(t)\rangle +\frac{%
A\varrho _{ul}^{(0)}}{2}\left[ \langle \hat{c}_{1}^{\dag }(t)\hat{c}%
_{1}(t)\rangle -\langle \hat{c}_{2}^{\dag }(t)\hat{c}_{2}(t)\rangle +1\right]
,  \label{e9} \\
\partial _{t}\langle \hat{c}_{1}(t)\hat{c}_{2}^{\dag }(t)\rangle &=&-\frac{%
\Gamma _{1}+\Gamma _{2}}{2}\langle \hat{c}_{1}(t)\hat{c}_{2}^{\dag
}(t)\rangle +\frac{A\varrho _{ul}^{(0)}}{2}\left[ \langle \hat{c}%
_{1}^{2}(t)\rangle -\langle \hat{c}_{2}^{\dag 2}(t)\rangle \right] ,
\label{e10}
\end{eqnarray}%
where $\Gamma _{1}=\kappa -A\varrho _{uu}^{(0)}$ and $\Gamma _{2}=\kappa
+A\varrho _{ll}^{(0)}.$ Here, we introduce the population inversion $\eta $
defined as $\varrho _{uu}^{(0)}=(1-\eta )/2$ with $-1\leqslant \eta
\leqslant 1$ \cite{Fesseha}. In addition, knowing that $\varrho
_{uu}^{(0)}+\varrho _{ll}^{(0)}=1$ and $|\varrho _{ul}^{(0)}|=\sqrt{\varrho
_{uu}^{(0)}\varrho _{ll}^{(0)}}$, one gets $\varrho _{ll}^{(0)}=$ $(1+\eta
)/2$ and $\varrho _{ul}^{(0)}=\varrho _{lu}^{(0)}=\sqrt{1-\eta ^{2}}/2$. By
setting $\partial _{t}\langle .\rangle =0$ in Eqs. [(\ref%
{e6})-(\ref{e10})], one can obtain the following solutions
\begin{eqnarray}
\langle \hat{c}_{1}\rangle _{ss} &=&\langle \hat{c}_{2}\rangle _{ss}=0
\label{e11} \\
\langle \hat{c}_{1}^{2}\rangle _{ss} &=&\langle \hat{c}_{2}^{2}\rangle
_{ss}=\langle \hat{c}_{1}\hat{c}_{2}^{\dag }\rangle _{ss}=0  \label{e12} \\
\langle \hat{c}_{1}^{\dag }\hat{c}_{1}\rangle _{ss} &=&\frac{-A\left( 1-\eta
\right) ^{2}}{4\left( \kappa +A\eta \right) \eta }+\frac{A\left( 1-\eta
^{2}\right) }{2\left( 2\kappa +A\eta \right) \eta },  \label{e13} \\
\langle \hat{c}_{2}^{\dag }\hat{c}_{2}\rangle _{ss} &=&\frac{-A\left( 1-\eta
^{2}\right) }{4\left( \kappa +A\eta \right) \eta }+\frac{A\left( 1-\eta
^{2}\right) }{2\left( 2\kappa +A\eta \right) \eta },  \label{e14} \\
\langle \hat{c}_{1}\hat{c}_{2}\rangle _{ss} &=&\frac{-A\left( 1-\eta \right)
\sqrt{1-\eta ^{2}}}{4\left( \kappa +A\eta \right) \eta }+\frac{A\sqrt{1-\eta
^{2}}}{2\left( 2\kappa +A\eta \right) \eta },  \label{e15}
\end{eqnarray}%
where $ss$ stands for steady-state. The Eqs. [(\ref{e13})-(\ref{e15})] are
physically meaningful only for $\eta \geqslant 0$, which imposes $0\leqslant
\eta \leqslant 1$.

Since the two laser modes $c_{1}$ and $c_{2}$ are shown to evolve in a
two-mode Gaussian state \cite{Hua Tan}, then, they can be characterized by
their $4\times 4$ covariance matrix $\sigma _{c_{1}c_{2}}$ defined by $\left[
\sigma _{c_{1}c_{2}}\right] _{kk^{\prime }}=\left( \langle \lbrack \hat{\nu}%
_{k},\hat{\nu}_{k^{\prime }}]_{+}\rangle \right) /2-\langle \hat{\nu}%
_{k}\rangle \langle \hat{\nu}_{k^{\prime }}\rangle $ (for $k$, $k^{\prime
}=\{1,..,4\}$) \cite{adesso}, where $\hat{\nu}^{\text{\textrm{T}}}=(\hat{x}%
_{1},\hat{y}_{1},\hat{x}_{2},\hat{y}_{2})$ is the vector of the position $%
\hat{x}_{j}=(\hat{c}_{j}^{\dag }+\hat{c}_{j})$ and momentum $\hat{y}_{j}=%
\mathrm{i}(\hat{c}_{j}^{\dag }-\hat{c}_{j})$ quadrature operators. Finally,
on the basis of Eqs. [(\ref{e13})-(\ref{e15})], the covariance matrix $%
\sigma _{c_{1}c_{2}}$ can be written in the following block form
\begin{equation}
\sigma _{c_{1}c_{2}}=\left(
\begin{array}{cc}
\sigma _{c_{1}} & \sigma _{c_{1}/c_{2}} \\
\sigma _{c_{1}/c_{2}}^{\mathrm{T}} & \sigma _{c_{2}}%
\end{array}%
\right) ,  \label{e16}
\end{equation}%
where the matrix $\sigma _{c_{1}}=\alpha _{11}$%
\mbox{$1 \hspace{-1.0mm}
{\bf l}$}$_{2}$($\sigma _{c_{2}}=\alpha _{22}$%
\mbox{$1 \hspace{-1.0mm}
{\bf l}$}$_{2}$) represents the laser mode $c_{1}$($c_{2}$), while $\sigma
_{c_{1}/c_{2}}=\beta _{12}\mathrm{diag}(1,-1)$ describes the correlations
between these two laser modes, with $\alpha _{jj}=2\langle \hat{c}_{j}^{\dag
}\hat{c}_{j}\rangle _{ss}+1$ for $j=1,2$, and $\beta _{12}=2\langle \hat{c}%
_{1}\hat{c}_{2}\rangle _{ss}$.

\section{Gaussian quantum steering of the two laser modes}

In the Ref. \cite{wiseman}, it has been shown that an arbitrary bipartite
Gaussian state $\hat{\varrho}_{c_{1}c_{2}}$ with covariance matrix $\sigma
_{c_{1}c_{2}}$ (\ref{e16}) is steerable (saying) from $c_{1}\rightarrow
c_{2} $, under Gaussian measurements, if the constraint

\begin{equation}
\sigma _{c_{1}c_{2}}+\mathrm{i}(0_{c_{1}}\oplus \Pi _{c_{2}})\geqslant \mathbf{0},  \label{e17}
\end{equation}%
is violated \cite{wiseman}, where $0_{c_{1}}$ being a $2\times 2$ null
matrix, and $\Pi _{c_{2}}$ $=\left(
\begin{array}{cc}
0 & 1 \\
-1 & 0%
\end{array}%
\right) $ is the symplectic matrix of the mode $c_{2}$ \cite{kogias}.

From an operational point of view, Kogias \textit{et al.} \cite{kogias} have
proposed a computable measure to quantify the amount by which a bipartite
Gaussian state $\hat{\varrho}_{c_{1}c_{2}}$ with covariance matrix $\sigma
_{c_{1}c_{2}}$ (\ref{e16}) is steerable, under Gaussian measurements, from $%
c_{1}\rightarrow c_{2}$. It is given by
\begin{equation}
\mathcal{G}^{c_{1}\rightarrow c_{2}}:=\mathrm{Max}[0,-\ln (\mu ^{c_{2}})],
\label{e18}
\end{equation}%
where $\mu ^{c_{2}}=\sqrt{\det \left( \sigma _{c_{2}}-\sigma _{c_{1}/c_{2}}^{%
\mathrm{T}}\sigma _{c_{1}}^{-1}\sigma _{c_{1}/c_{2}}\right) }$ is the
symplectic eigenvalue of the Schur complement \cite{kogias} of the mode $%
c_{1}$ in the covariance matrix (\ref{e16}). The measure $\mathcal{G}%
^{c_{1}\rightarrow c_{2}}$ vanishes when the state $\hat{\varrho}%
_{c_{1}c_{2}}$ is nonsteerable from $c_{1}\rightarrow c_{2}$.

Particularly, when the steered party is a single mode, Eq. (\ref{e18}) takes
the following simple expression \cite{kogias}%
\begin{equation}
\mathcal{G}^{c_{1}\rightarrow c_{2}}=\mathrm{Max}[0,\frac{1}{2}\ln \frac{%
\det \sigma _{c_{1}}}{\det \sigma _{c_{1}c_{2}}}],  \label{e19}
\end{equation}%
where the steering in the reverse direction can be obtained by changing the
role of the modes $c_{1}$ and $c_{2}$ in Eq. (\ref{e19}), i.e., $\mathcal{G}%
^{c_{2}\rightarrow c_{1}}=\mathrm{Max}[0,\frac{1}{2}\ln \frac{\det \sigma
_{c_{2}}}{\det \sigma _{c_{1}c_{2}}}].$ We notice that a non zero value of $%
\mathcal{G}^{c_{1}\rightarrow c_{2}}$($\mathcal{G}^{c_{2}\rightarrow c_{1}}$%
) implies that the state $\hat{\varrho}_{c_{1}c_{2}}$ is steerable from $%
c_{1}\rightarrow c_{2}$($c_{2}\rightarrow c_{1}$), and three different cases
can be observed: (\textit{1}) two-way steering, where $\mathcal{G}%
^{c_{1}\rightarrow c_{2}}>0$ and $\mathcal{G}^{c_{2}\rightarrow c_{1}}>0$, (%
\textit{2}) no-way steering, where $\mathcal{G}^{c_{1}\rightarrow c_{2}}=%
\mathcal{G}^{c_{2}\rightarrow c_{1}}=0$, and (\textit{3}) one-way steering,
in which the state $\hat{\varrho}_{c_{1}c_{2}}$ is steerable only in one
direction: $\mathcal{G}^{c_{2}\rightarrow c_{1}}=0$ with $\mathcal{G}%
^{c_{1}\rightarrow c_{2}}>0$ or $\mathcal{G}^{c_{1}\rightarrow c_{2}}=0$
with $\mathcal{G}^{c_{2}\rightarrow c_{1}}>0$.

Furthermore, we compare the steering of the two modes $c_{1}$ and $c_{2}$
with their quantum entanglement. In this respect, to quantify the
entanglement of the laser modes $c_{1}$ and $c_{2}$, we use the Gaussian R%
\'{e}nyi-2 entanglement $\mathcal{E}_{2}$ defined---for a two-mode squeezed
thermal state \cite{sts} with covariance matrix $\sigma _{c_{1}c_{2}}$ (\ref%
{e16})---as \cite{R2,adesso}%
\begin{equation}
\mathcal{E}_{2}=\left\{
\begin{array}{c}
\frac{1}{2}\ln \left[ \frac{\left[ (g+1)s_{+}-\sqrt{\left[
(g-1)^{2}-4s_{-}^{2}\right] \left[ s_{+}^{2}-s_{-}^{2}-g\right] }\right] ^{2}%
}{4(s_{-}^{2}+g)^{2}}\right] \text{ if }2|s_{-}|+1\leqslant g<2s_{+}-1, \\
0\text{ \ \ \ \ \ \ \ \ \ \ \ \ \ \ \ \ \ \ \ \ \ \ \ \ \ \ \ \ \ \ \ \ \ \
\ \ \ \ \ \ \ \ \ \ \ \ \ \ \ \ \ \ if \ }g\geqslant 2s_{+}-1,%
\end{array}%
\right.  \label{e20}
\end{equation}%
where $s_{\pm }=(\alpha _{11}\pm \alpha _{22})/2$ and $g=\left( \alpha
_{11}\alpha _{22}-\beta _{12}^{2}\right) $.

We notice that the Gaussian R\'{e}nyi-2 entanglement $\mathcal{E}_{2}$ does
not increase under Gaussian local operations and classical communication, it
is additive under tensor product states, and it is monogamous \cite{R2}.

\begin{figure}[th]
\centerline{\includegraphics[width=0.45\columnwidth,height=5.5cm]{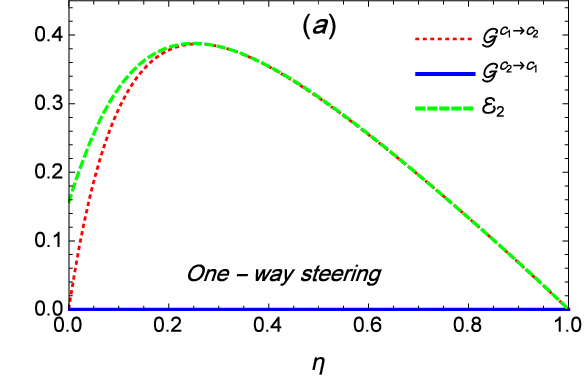}
		\includegraphics[width=0.45\columnwidth,height=5.5cm]{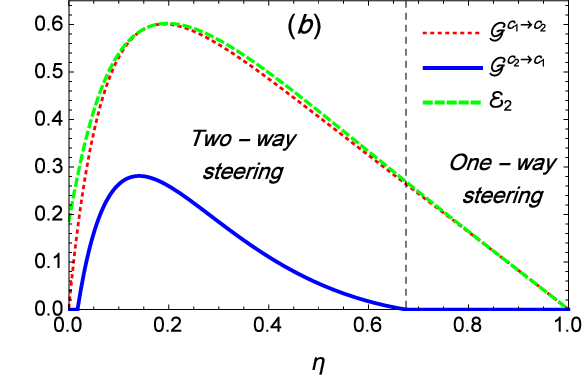}}
\centerline{\includegraphics[width=0.45\columnwidth,height=5.5cm]{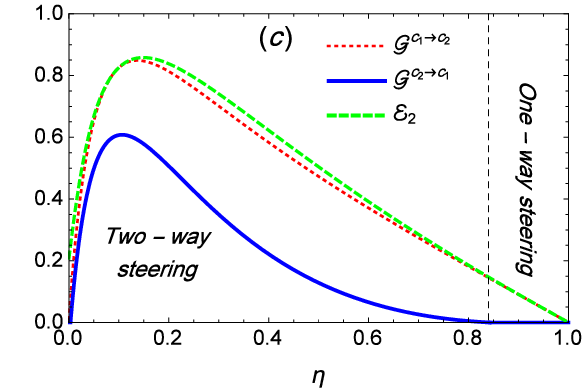}
		\includegraphics[width=0.45\columnwidth,height=5.5cm]{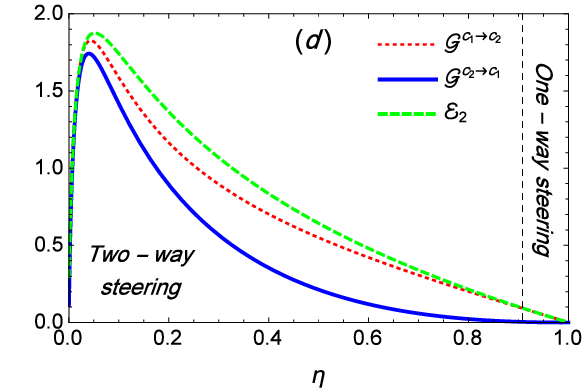}}
\caption{The Gaussian steering $\mathcal{G}^{c_{1}\rightarrow c_{2}}$, $%
\mathcal{G}^{c_{2}\rightarrow c_{1}}$ and the Gaussian R\'{e}nyi-2
entanglement $\mathcal{E}_{2}$ as functions of the population inversion $%
\protect\eta $ using various values of the linear gain coefficient: (a) $%
A=50~\text{kHz}$, (b) $A=100~\text{kHz}$, (c) $A=200~\text{kHz}$, and (d) $%
A=2000~\text{kHz}$. The cavity decay rate is fixed as $\protect\kappa=3.85~%
\text{kHz}$. Panel (a) shows an interesting situation in which the state $%
\hat{\protect\varrho}_{c_{1}c_{2}}$ is entangled, however, it is steerable
only from $c_{1} \rightarrow c_{2}$, which clearly reflects the asymmetric
property of quantum correlations. Quite remarkably, panel (d) shows that
maximum steering and entanglement can be achieved by inducing maximum quantum coherence in the cavity, which corresponds to $%
\protect\eta\rightarrow 0 $ and $A\gg1$.}
\label{f2}
\end{figure}

For realistic estimation of both steering and entanglement of the two laser
modes $c_{1}$ and $c_{2}$, we use experimental parameters from \cite%
{Han,Meschede}: the cavity decay rate for the laser modes c$_{1}$ and c$_{2}$
is $\kappa =3.85~\text{kHz}$, the atomic damping rate $\gamma =20~\text{kHz}
$, the atomic coupling strength $\epsilon=43~\text{kHz,}$ and $\rho =22~%
\text{kHz}$ as value of the rate at which the atoms are initially injected
into the cavity. Using these parameters, one has $A=2\rho \epsilon
^{2}/\gamma ^{2}\approx 200~\text{kHz}$ for the linear gain coefficient.

In Fig. \ref{f2}, we plot the steerabilities $\mathcal{G}^{c_{1}\rightarrow
c_{2}}$ and $\mathcal{G}^{c_{2}\rightarrow c_{1}}$ as well as the Gaussian R%
\'{e}nyi-2 entanglement $\mathcal{E}_{2}$ as functions of the population
inversion $\eta $ using various values of the linear gain coefficient $A$.
First, we remark that the steering and entanglement behave similarly under
the same circumstances. In addition, the values of the steering $\mathcal{G}%
^{c_{1}\rightarrow c_{2}}$ are fully differ from those of the steering in
the reverse direction $\mathcal{G}^{c_{2}\rightarrow c_{1}}$, which clearly
reflects the role of the measurement in quantum mechanics. Interestingly,
Fig. \ref{f2} shows that for $\eta =1$, the steering in both directions $%
c_{1}\rightleftarrows c_{2}$ as well as the entanglement $\mathcal{E}_{2}$
vanish regardless of the choice of the linear gain coefficient $A$. These
results can be explained as follows: since $\eta =1$ corresponds to the case
in which all the atoms are initially prepared in the lower energy level $%
|l\rangle $, then no possibility for radiation emission by the atoms in this
case, and further no quantum correlations (including steering and
entanglement) can be generated between the modes $c_{1}$ and $c_{2}$.
Besides, when $\eta \rightarrow 0$, which corresponds to the situation in
which the atoms are initially prepared with maximum quantum
coherence, we remark that maximum steering and entanglement could be
achieved. Strikingly, Fig. \ref{f2}(a) shows that the state $\hat{\varrho}%
_{c_{1}c_{2}}$ is steerable only from $c_{1}\rightarrow c_{2}$ (one-way
steering), although the two modes $c_{1}$ and $c_{2}$ are entangled. This
could be explained as follows: Alice (owning the mode $c_{1}$) and Bob
(owning the mode $c_{2}$) can implement the same Gaussian measurements on
their shared entangled state $\hat{\varrho}_{c_{1}c_{2}}$, however obtain
different results, which reflects the asymmetric property of quantum
correlations. In other words, one-way steering from $c_{1}\rightarrow c_{2}$
means that Alice can convince Bob that their shared state is entangled, in
contrast, the reverse process is impossible. This is partly due to the
asymmetric form of the covariance matrix (\ref{e16}), wherein $%
\sigma_{1}\neq\sigma_{2}$, and partly due to the definition of the
steerabilities $\mathcal{G}^{c_{1}\rightarrow c_{2}}$ and $\mathcal{G}%
^{c_{2}\rightarrow c_{1}}$.

In particular, Figs. \ref{f2}(b), \ref{f2}(c) and \ref{f2}(d) show that by
using high values of the linear gain coefficient $A$, two-way steering can
be observed, and further it can be transformed into one-way steering when $%
\eta \rightarrow 1.$%
\begin{figure}[h]
\centerline{\includegraphics[width=0.45\columnwidth,height=5.5cm]{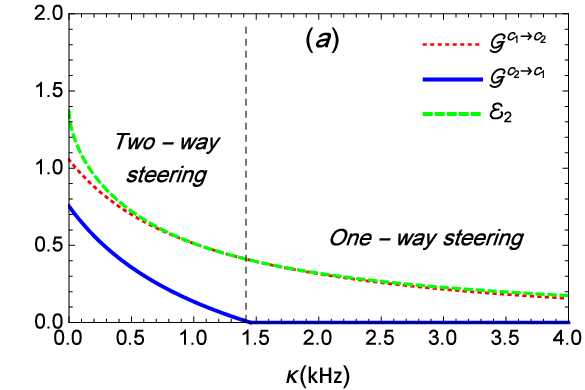}%
\includegraphics[width=0.45\columnwidth,height=5.5cm]{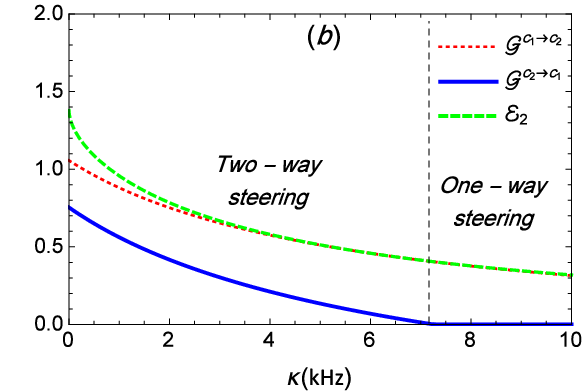}}
\caption{The Gaussian steering $\mathcal{G}^{c_{1}\rightarrow c_{2}}$, $%
\mathcal{G}^{c_{2}\rightarrow c_{1}}$ and the Gaussian R\'{e}nyi-2
entanglement $\mathcal{E}_{2}$ against the cavity decay rate $\protect\kappa
$ using two different values of the linear gain coefficient: (a) $A=20~\text{%
kHz}$, and (b) $A=100~\text{kHz}$. The population inversion is fixed as $%
\protect\eta =0.25$. As expected, both steering and entanglement decay under
the cavity losses, and they can be enhanced by augmenting the coefficient $A$%
, i.e., the rate at which the atoms are injected into the cavity. Similarly
to Fig. \protect\ref{f2}, Fig. \protect\ref{f3} shows that one-way steering
still occurs only from $c_{1}\rightarrow c_{2}$ and the steering $\mathcal{G}%
^{c_{1}\rightarrow c_{2}}$ and $\mathcal{G}^{c_{2}\rightarrow c_{1}}$ remain
less than entanglement $\mathcal{E}_{2}$. }
\label{f3}
\end{figure}
\newline

\begin{figure}[h]
\centerline{\includegraphics[width=0.4\columnwidth,height=5cm]{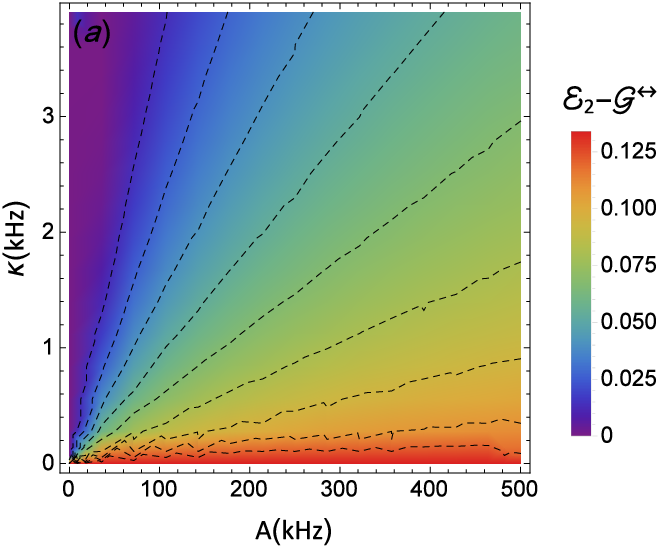}%
\includegraphics[width=0.4\columnwidth,height=5cm]{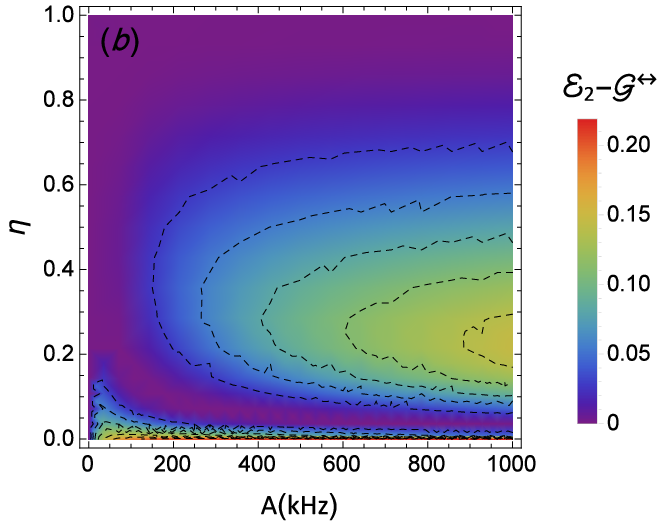}}
\caption{(a) Density plot of the difference $\mathcal{E}_{2}-\mathcal{G}%
^{\leftrightarrow }$ (with $\mathcal{G}^{\leftrightarrow }=\text{Max}[%
\mathcal{G}^{c_{1}\rightarrow c_{2}}$, $\mathcal{G}^{c_{2}\rightarrow c_{1}}]
$) versus $\protect\kappa $ and $A$ using $\protect\eta =0.5$. (b) Density
plot of $\mathcal{E}_{2}-\mathcal{G}^{\leftrightarrow }$ versus $\protect%
\eta $ and $A$ using $\protect\kappa =3.85~\text{kHz}$. In various
circumstances, we remark that the Gaussian R\'{e}nyi-2 entanglement $%
\mathcal{E}_{2}$ is always greater or equal to the maximum steering $%
\mathcal{G}^{\leftrightarrow }$, which is consistent with the Ref.
\protect\cite{kogias}.}
\label{f4}
\end{figure}

On the other hand, Fig. \ref{f3} illustrates the dependence of the steering $%
\mathcal{G}^{c_{1}\rightarrow c_{2}}$ and $\mathcal{G}^{c_{2}\rightarrow
c_{1}}$ and the Gaussian R\'{e}nyi-2 entanglement $\mathcal{E}_{2}$ on the
cavity decay rate $\kappa $ using $A=20~$kHz in panel (a), and $A=100~$kHz
in panel (b). The population inversion is fixed as $\eta =0.25$. As shown, $%
\mathcal{G}^{c_{1}\rightarrow c_{2}}$, $\mathcal{G}^{c_{2}\rightarrow c_{1}}$
and $\mathcal{E}_{2}$ are maximum for $\kappa=0$, and have a tendency to
diminish with increasing dissipation $\kappa$. Similarly to the results
presented in Fig. \ref{f2}, Fig. \ref{f3} shows that one-way steering still
occurs only from $c_{1}\rightarrow c_{2}$, which emerges from the fact that
the steering $\mathcal{G}^{c_{1}\rightarrow c_{2}}$ remains greater than the
steering $\mathcal{G}^{c_{2}\rightarrow c_{1}}$. Importantly, Figs. \ref{f2}
and \ref{f3} show together that the range of one-way steering is strongly
influenced by the values of linear gain coefficient $A$. Additionally, they
show that the steering $\mathcal{G}^{c_{1}\rightarrow c_{2}}$ and $\mathcal{G%
}^{c_{2}\rightarrow c_{1}}$ are upper bounded by the Gaussian R\'{e}nyi-2
entanglement $\mathcal{E}_{2}$, which is well supported by the results
illustrated in Fig. \ref{f4}. Indeed, with density values of the physical
parameters $A$, $\kappa $ and $\eta $ of the state $\hat{\varrho}%
_{c_{1}c_{2}}$, Fig. \ref{f4} shows that the difference $\mathcal{E}_{2}-%
\mathcal{G}^{\leftrightarrow}$, where $\mathcal{G}^{\leftrightarrow}=\mathrm{%
Max[\mathcal{G}^{c_{1}\rightarrow c_{2}},\mathcal{G}^{c_{2}\rightarrow
c_{1}}]}$ \cite{kogias}, remains always positive or equal to zero, meaning
that the maximum steering $\mathcal{G}^{\leftrightarrow}$ cannot exceed the
Gaussian R\'{e}nyi-2 entanglement $\mathcal{E}_{2}$ of the two laser modes $%
c_{1}$ and $c_{2}$. This therefore is consistent with the Ref. \cite{kogias}.

Quite remarkably, Figs. \ref{f2} and \ref{f3} show that one-way steering is
occurred---within various circumstances---only in the direction $%
c_{1}\rightarrow c_{2}$. In what follows, we demonstrate that---in
general---the state $\hat{\varrho}_{c_{1}c_{2}}$ can exhibit one-way
steering behavior only from $c_{1}\rightarrow c_{2}$.
Indeed, since $\mathcal{G}^{c_{1}\rightarrow c_{2}}\geqslant \mathcal{G}%
^{c_{2}\rightarrow c_{1}}$ (which includes $\mathcal{G}^{c_{1}\rightarrow
c_{2}}>0$ and $\mathcal{G}^{c_{2}\rightarrow c_{1}}=0$) is a necessary
condition for observing one-way steering from $c_{1}\rightarrow c_{2}$,
hence, the general fulfilling of the constraint $\mathcal{G}%
^{c_{1}\rightarrow c_{2}}\geqslant \mathcal{G}^{c_{2}\rightarrow c_{1}}$
implies that one-way steering can occur only from $c_{1}\rightarrow c_{2}$.
Then, utilizing Eqs. (\ref{e16}) and (\ref{e19}), one can show that $%
\mathcal{G}^{c_{1}\rightarrow c_{2}}\geqslant \mathcal{G}^{c_{2}\rightarrow
c_{1}}$ is equivalent to $\langle \hat{c}_{1}^{\dag }\hat{c}_{1}\rangle
_{ss}\geqslant \langle \hat{c}_{2}^{\dag }\hat{c}_{2}\rangle _{ss}$, where $%
\langle \hat{c}_{1}^{\dag }\hat{c}_{1}\rangle _{ss}$($\langle \hat{c}%
_{2}^{\dag }\hat{c}_{2}\rangle _{ss}$) is the mean photon number in the
laser mode $c_{1}$($c_{2}$). Moreover, employing Eqs. (\ref{e13}) and (\ref%
{e14}), we obtain $\langle \hat{c}_{1}^{\dag }\hat{c}_{1}\rangle
_{ss}-\langle \hat{c}_{2}^{\dag }\hat{c}_{2}\rangle _{ss}=\frac{A\left(
1-\eta \right) }{2\left( \kappa +A\eta \right) }$, which is always positive
because $0\leqslant \eta \leqslant 1$, $A\geqslant 0$, and $\kappa \geqslant
0$. This means that the condition $\mathcal{G}^{c_{1}\rightarrow
c_{2}}\geqslant \mathcal{G}^{c_{2}\rightarrow c_{1}}$ is always satisfied. Therefore, one-way steering can occur between the modes $c_{1}$ and $c_{2}$ only in the direction $c_{1}\rightarrow c_{2}$. The
demonstrated unidirectional one-way steering may lead to more consideration
in the application of asymmetric quantum steering \cite{Uola}.

Finally, with the availability of cavity-quantum electrodynamics and trapped
atoms techniques \cite{Rempe,QED}, and the homodyne detection method
developed in \cite{Laurat}, these results may be verified experimentally.

\section{Conclusion}

In a two-mode Gaussian state $\hat{\varrho}_{c_{1}c_{2}}$, asymmetric
quantum steering is studied. The modes $c_{1}$ and $c_{2}$ are generated,
respectively, during the first and second transition of a nondegenerate
three-level laser. By means of the master equation that governs the dynamics
of the state $\hat{\varrho}_{c_{1}c_{2}}$, we derived analytical expression
of the stationary covariance matrix describing any Gaussian state of the two
laser modes $c_{1}$ and $c_{2}$ . To quantify the amount by which the state $%
\hat{\varrho}_{c_{1}c_{2}}$ is steerable in a given direction, we employed
the measure proposed in \cite{kogias}. Using realistic experimental
parameters, we showed that it is possible to generate stationary Gaussian
steering in both directions $c_{1}\rightleftarrows c_{2}$ (two-way
steering), and even one-way steering. Also, we showed that by an appropriate
choice of the physical parameters of the state $\hat{\varrho}_{c_{1}c_{2}}$ (quantum coherence and cavity decay rate), the region of
one-way steering can be controlled, which may have potential applications in
asymmetric quantum information tasks \cite{Uola}. Importantly, we
demonstrated that in the considered state $\hat{\varrho}_{c_{1}c_{2}}$: (%
\textit{i}) the $c_{2}\rightarrow c_{1}$ steerability cannot exceed that in
the reverse direction, (\textit{ii}) one-way steering can, in general, occur
only from $c_{1}\rightarrow c_{2}.$ On the other hand, we compared the
Gaussian steering of the two laser modes $c_{1}$ and $c_{2}$ with their
corresponding entanglement quantified by means of the Gaussian R\'{e}nyi-2
entanglement. These two kind of quantum correlations (steering and
entanglement) are found to behave similarly under the same circumstances,
i.e., they are found to decay under influence of the cavity losses, while
they could be well enhanced by inducing more and more quantum coherence into the cavity. In particular, it is
found that the steering remains upper bounded by the Gaussian R\'{e}nyi-2
entanglement. With these results, we believe that nondegenerate three-level
laser systems may have immediate applications in the implementing of
asymmetric quantum information processing and communication.

\end{document}